\pdfoutput=1
\documentclass[aps,twocolumn, floatfix, prb]{revtex4}
\usepackage{graphicx}
\DeclareGraphicsExtensions{.pdf}
\usepackage{amsmath,amssymb,bbold,bm}
\usepackage{float}
\usepackage{epstopdf}

\newcommand{\bk}{{\bm k}}
\newcommand{\bw}{{\bm w}}

\newcommand{\bR}{{\bm R}}

\newcommand{\bG}{{\bm G}}
\newcommand{\bK}{{\bm K}}

\newcommand{\blam}{{\bm \lambda}}

\newcommand{\bp}{{\bm p}}
\newcommand{\bq}{{\bm q}}
\newcommand{\bb}{{\bm b}}

\begin{document}

\title{Weyl Semimetal from the Honeycomb Array of Topological Insulator Nanowires}

\author{M. M. Vazifeh}
\affiliation{Department of Physics and Astronomy, University of
British Columbia, Vancouver, BC, Canada V6T 1Z1}

\begin{abstract}
We show that isolated Weyl nodes can arise in a system of parallel topological insulator nano-wires arranged in a honeycomb fashion. 
This introduces another theoretical example of a topological semi-metal phase with more than one pair of Weyl nodes 
and due to the simple form of its Hamiltonian it can be used to study various interesting phenomena associated with this phase.
Our result emphasizes that depending on the separation of the Weyl nodes a topological electromagnetic response might 
or might not emerge and two pairs might have overall cancelling contributions to the net anomalous Hall conductivity.   
\end{abstract}

\date{\today}

\maketitle

\section{Introduction}

New topological phases of matter have been discovered recently highlighting the fact that the topological structures formed by the underlying electronic states are as important \cite{Wenbook} as the symmetries involved and can be closely related to some interesting physically observable phenomena \cite{moore1, hasan_rev, hasan1, hasan2, shen1, hasan3, hasan4, zhangprb, qi2, wormhole, tse1}. A recently discovered example in two and three dimensions is the time-reversal invariant topological band insulator phase which can be formed by the electronic states in crystals with strong spin-orbit couplings and has been fully classified in terms of well-defined topological invariants \cite{topoClass1,fu-kane3D}. In a strong topological insulator (TI) phase, the quantum states have been formed in a non-trivial fashion so that the system belongs to a different topological class than the ordinary insulator (OI) meaning that there is no fully gapped deformation path in the phase space of the Hamiltonians that connects the Hamiltonian of the system to that of an ordinary insulator preserving the time-reversal symmetry \cite{murakami1}. Any phase space path connecting two topologically distinct points in the phase space would contain a point where the associated Hamiltonian is gapless. This intuitively explains the presence of the robust topological surface states at the interfaces between these systems and ordinary insulators. It can be understood from the fact that one can establish a mapping between any path connecting a point in the bulk of a topological insulator to a point in the vacuum and the phase space paths which respect time-reversal symmetry connecting two topologically distinct regions and therefore it must contain a critical point where the system is locally gapless and all such points span the metallic surface. Therefore, any local perturbation on the surface that respects the time-reversal symmetry cannot destroy the gapless surface modes. These gapless surface modes are then protected by the coexistence of time-reversal symmetry on the surface and the non-trivial topological structure in the bulk of the system. 

%
The topological protection is not restricted to just one and two dimensional metallic systems. Indeed one can generalize this idea to find a three dimensional metallic system where the band crossing is protected by the underlying nontrivial topological structure of the quantum states rather than discrete symmetries such as the inversion or the time-reversal. Recently, experimentally feasible proposals to realize such 3D metallic phases has been made where energy crossings (Dirac Points) exist in the band dispersion and are protected by global topological properties of the electronic states. \cite{weyl1, weyl2, weyl3, weyl4, weyl5} In this case there is no discrete symmetry involved in protecting the Dirac point from the gap opening as opposed to the 2D case where we still need to keep the time-reversal symmetry in order to maintain the band crossing  \cite{gap1,gap2}. The existence of this nontrivial topological structure associated with the isolated energy crossings would lead to some interesting transport phenomena in the bulk and on the surface of such systems. A topological axion term with a non-uniform axion field emerges in the effective electromagnetic Lagrangian which explains the chiral anomaly in these systems \cite{anomaly1,anomaly2,anomaly3, anomaly4}    

%
In general, accidental energy crossings at certain points in the Brillouin zone (BZ) of a three dimensional electronic system can only exist by fine tuning and imposing certain symmetries in a model. These are often hard to achieve and symmetries are sometimes broken in actual materials due to the imperfections or interactions. However, recently it has been suggested that in certain systems energy crossings can have topological protection. When such topological protection exists, the system is in a topological semi-metal phase even in the presence of weak disorder or other types of interactions that preserve the momentum conservation. These band crossing points in the BZ can only exist in pairs \cite{Nielsen} and are called Weyl points. Near an isolated Weyl point the low lying states can be described by a 2-by-2 Dirac Hamiltonian. Therefore, any local perturbation that does not violate the momentum conservation can only shift the position of such a point in the BZ and is not sufficient to gap out the spectrum unless it is strong enough to merge two such points with opposite chirality and make the system unstable towards becoming an insulator. These points can be thought of as topological defects in the fibre bundle formed by the electronic states in the BZ as the base manifold and are the magnetic monopoles of the pseudo-magnetic field associated with the gauge field defined for the Bloch states. Only the pairwise annihilation of such points with opposite chirality is possible as one can define a well-defined charge for these monopoles which is a conserved quantity in the regime where the crystal momentum is still a good quantum number.

%
The question is how such energy crossings can occur in a band system. Theoretically, it is possible to choose a physical parameter to adiabatically drive a system from a topological insulator phase to an ordinary insulator or vice versa by tuning such a parameter. As one fine tunes such a parameter to the critical point where the gap closes (which is inevitable when we have time-reversal symmetry) the system becomes an unstable bulk metal if the time-reversal and inversion symmetry is preserved at the same time. At this critical point the system would be in an unstable 3D metallic phase with degenerate bands near the crossing points. This gapless metallic phase is susceptible to local perturbations and instabilities that drive the system to either topological or ordinary insulator phases. Now if another parameter in the Hamiltonian can be tuned in such a way that it separates the degenerate bands at the crossing points in the BZ (by breaking the inversion and/or the time-reversal symmetry) then the system would be in a topological semi-metal phase with gapless modes dispersing in three spatial dimensions which are "robust" against momentum conserving perturbations. The topological nature of the phase reflects itself in the appearance of a nontrivial term in the effective Lagrangian of the electromagnetic fields \cite{anomaly4}. This topological term is an abelian axion term with a non-homogeneous axion field

\begin{equation}
S_{\text{axion}} = C \int  \theta(x) \mathcal{F} \wedge \mathcal{F},
\end{equation}
where C is a constant and $\theta(x)$ modulates in space and time with wave-vectors that depends on the separation of the Weyl nodes in momentum and energy and for a single pair of Weyl nodes it is given by $\theta(x) = q \cdot x$ in which $q$ is a four-vector of the separation of the pair of the Weyl nodes in the crystal momentum and energy \cite{anomaly1,anomaly2,anomaly3}. This topological term in the effective electromagnetic Lagrangian is closely related to the topological transport phenomena that is present in these systems\cite{anomaly4}. It turns out that a system with more than one pairs of Weyl fermions might have zero net anomalous Hall conductivity since pairs can have total cancelling contributions due to the exactly opposite separations of the them in momentum space. Here we consider a simple theoretical model which can lead to a topological semi-metal phase with two Weyl pairs under certain circumstances. Although there are big challenges on the way of the experimental realization of such a system, it can be used as a platform to study the bulk and the surface properties of a system in a Weyl semi-metal phase when there are more than one pair of Weyl nodes.    

The paper is organized as follows: In section II, we provide a simple mathematical proof of the topological protection of the Weyl nodes using the Clifford algebra when we have a pair of them separated in the BZ. In section III, we introduce a lattice model made by arranging parallel TI nano wires in a honeycomb fashion with a lattice spacing which is small enough to allow a significant hopping of the electrons between these wires and then we discuss how one can achieve a topological semimetal phase out of this arrangement.

\section{Topological Semi-metal}

At low energies, the minimal Hamiltonian describing the unstable 3D gapless system at the TI-OI transition point with a pair of energy crossing points would have the following form

\begin{equation}
\mathcal{H} = \sum_{\alpha=\pm} \sum_{ |\bk - \bw_{ \alpha}| < \Lambda}  \Psi^{\dag}_{\alpha} ( \bk)[ \blam_{ \alpha}.(\bk - \bw_{\alpha})] \Psi^{}_{\alpha}(\bk),
\end{equation}
where $\blam_{\alpha}$ are 2-by-2 matrices which are given in terms of Pauli matrices, $\sigma_i$, by 

\begin{equation}
\blam = (\upsilon_{ x} \sigma_x, \upsilon_{y} \sigma_y, \upsilon_{z} \sigma_z),
\end{equation} 
$\bw_{+} = \bw_{-} = \bw_0$ are the position vectors of the Weyl points with $+1$ and $-1$ chirality ( $=$ sgn($\upsilon_x \upsilon_y \upsilon_z$)). This Hamiltonian consists of doubly degenerate Dirac cones with the Dirac points located at $\bw_0$. The two component spinor operators $\Psi^{\dag}_{\pm}$ are associated with the modes on the Dirac cone with chirality $\alpha = \pm$. This Hamiltonian describes the low-energy gapless phase at the phase transition point between a topological insulator and an ordinary insulator. The cutoff, $\Lambda$, defines the momentum range around each Weyl point for which the linear approximation is valid. When the opposite chirality Weyl points happen at the same point in the BZ, there is no topological protection and even momentum conserving perturbations that connect the degenerate states can drive the system to an insulator phase. Now, if somehow the opposite chirality points become separated ($\bw_{+} - \bw_{-} = \bq \neq 0$) then the system would be in a gapless topological phase where small local momentum conserving perturbations cannot produce a gap in the spectrum and would only shift the position of the Weyl points in the BZ. 

Using the mathematical properties of the Clifford algebra, it is possible to see how such protection can occur for a gapless system containing a pair of separated opposite chirality Weyl points. The local Hamiltonian matrix in the BZ describing the low energy states in the topological semi-metal phase containing a pair of opposite chirality Weyl points can be written in terms of 4-by-4 Dirac matrices as follows 
\begin{equation}\label{hamil1}
{\mathcal H(\bk)} = k_x \gamma_1 + k_y \gamma_2 + k_z \gamma_3 + \frac{q}{2} \gamma_3',
\end{equation}
where for simplicity and without the loss of generality we have put $|\upsilon_{x,y,z}|=1$. The first three 4-by-4 matrices $\gamma_{1,2,3}$ are mutually anti-commuting to ensure that we will get two opposite chirality Dirac nodes dispersing in three dimensions. Without $\gamma_3'$, we get two degenerate Dirac cones, therefore we must introduce it into the Hamiltonian to separate the Weyl nodes by $q$ in momentum space and it must commute with at least one of the other matrices or a momentum independent linear combination of them depending on the direction of the desired separation. This commutation relation ensures that the effect of the added term is only opposite shifts of the Weyl nodes along an specific direction in the momentum space. We consider Weyl nodes separated in $k_z$ direction, therefore, we must choose $\gamma_3'$ in a way that it commutes with the matrix associated with $k_z$, i.e., $\gamma_3$ and anti-commutes with the rest. Note that by writing the Hamiltonian local in momentum space, we have implicitly excluded any perturbation that connects two Weyl nodes which might arise from the scattering sources. Those can potentially gap out the Weyl nodes by connecting states with different momentums, however, the irrational separation of the Weyl nodes ensures that the system is not unstable toward the charge density wave ordering with a commensurate wave vector.  To summarize, we have $\{\gamma_i,\gamma_j\}=\{\gamma_{1,2},\gamma_3'\}=0$ and $[\gamma_3, \gamma_3'] = 0$. Note that the low energy Hamiltonian describing any pair of opposite chirality Weyl nodes is connected by a well defined unitary transformation to the Hamiltonian given by Eq. (\ref{hamil1}). The spectrum of this Hamiltonian can be obtained by, first, multiplying the Hamiltonian by itself and then taking the square root of the resulting diagonal matrix. It is easy to check that the relation between the above matrices leads to the following spectrum for this Hamiltonian

\begin{equation}
\varepsilon_{\bk,s} = \pm \sqrt{k_x^2+k_y^2 + (k_z+  \frac{s q}{2})^2}, \;\;\;\;\;\;\; s=\pm 1,
\end{equation}

and it has all the correct properties, i.e., the opposite chirality Weyl nodes separated by q in the $k_z$ direction. If there exist a momentum conserving perturbation which opens up a gap in the spectrum of the above Hamiltonian, there must be an associated hermitian matrix that anti-commutes with all of the terms in ${\mathcal H(\bk)}$. If such a term does not anti-commute with all the terms, then it can only change the position of the Weyl nodes. This can be seen by following the same steps we have taken to obtain the spectrum of ${\mathcal H(\bk)}$. The statement we want to prove here is that \emph{no such term} exists and any 4-by-4 Hermitian matrix which can be written as a superposition of the normalized clifford algebra matrices, $ \gamma_g$, (${\gamma_g}^2 = {I}_{4 \times 4}$) commutes with at least one of the terms in this Hamiltonian. Therefore, it cannot produce a gap in the spectrum since this way it can only shift the positions of the Weyl nodes in momentum space. In order to prove this statement, we show that the maximum number of mutually anti-commuting matrices for which one of them commutes with $\gamma_3'$ is three. In order to do so we need to choose a representation of the present matrices. Since $\gamma_3$ and $\gamma_3'$ commute with each other, we can choose a basis in which both of these matrices are diagonal. This way we have $\gamma_3 = \text{diag}[1,a_1,a_2,-1-a_1-a_2]$ and $\gamma_3' = \text{diag}[1,b_1,b_2,-1-b_1-b_2]$. It turns out that only three independent choices can satisfy this commutation condition, i.e., $(a_1,a_2,b_1,b_2) = (1,-1,-1,1),(-1,1,1,-1),(-1,1,-1,-1)$. For each of these choices we can show that only two independent normalized matrices exist which can simultaneously anti-commute with $\gamma_3$ and $\gamma_3'$ and with themselves. This would then imply that the maximum possible number of mutually anti-commuting matrices, when we require that one of them to commute with $\gamma_3'$, is three . These three have already been exploited in the Hamiltonian by coupling to three momentum components and therefore no other matrices can be added to anti-commute with all of these matrices and $\gamma_3'$ in order to gap out the spectrum. To see how it works, consider the most general normalized matrix satisfying the aforementioned relations for the first choice of $(a_1,a_2,b_1,b_2)$. For this particular representation it has the following form

\begin{equation}
\gamma_p = \left( \begin{matrix}
0 & 0 & 0 & e^{i \phi} \\
0 & 0& e^{i \theta} & 0 \\
0 &e^{-i \theta} & 0 & 0 \\
e^{-i \phi} & 0 & 0 & 0
\end{matrix} \right),
\end{equation}

Matrices of this form that anti-commute with each other must satisfy $|\phi_i - \phi_j| = |\theta_i - \theta_j| = \pi/2$. It is easy to see that only two independent matrices can be found in this representation to satisfy these equations. These matrices are already exploited in the Hamiltonian by coupling to momentum components $k_x$ and $k_y$ and therefore no other matrices can be added to the Hamiltonian given in Eq. (\ref{hamil1}) that anti-commutes with all the present matrices. The same argument can be established for the other two choices of $(a_1,a_2,b_1,b_2)$. Therefore, Weyl nodes cannot be annihilated unless two of them with opposite chirality merge at the same point in the BZ. This cannot be possible without the sufficiently strong local perturbations ($ \sim  |\hbar \upsilon_F q|$ to bring two nodes to the same point in BZ) and the system is in a topological semimetal phase in at least a region of the phase space. The above proof is a generalization of the previously known argument in the literature which one considers only one of the isolated Weyl nodes in a two band model using two-by-two Pauli matrices. Here we have considered all the four bands which are present in the low energy theory describing a pair of Weyl fermions. In the rest of this paper we present a system with more than one pair of Weyl nodes where the stability arguments should be considered more carefully as it might be possible to connect two pairs of Weyl nodes with a charge density wave ordering.


\section{TI Nanowires}
The surface modes of a cylindrical topological insulator system can be described by a Dirac Hamiltonian in a curved space \cite{Curved} ($\hbar = 1$)

\begin{equation}\label{hcurve}
H_k = \frac{\upsilon}{2} [{\bm \nabla} \cdot \hat{\bm n} + \hat{\bm n} \cdot (\bp \times {\bm \sigma}) + (\bp \times {\bm \sigma}) \cdot \hat{\bm n}],
\end{equation}
where $\bm \sigma$ is the pauli matrix vector which acts on the spin Hilbert space. $\bp = -i \bm \nabla$ is the momentum operator and $\hat{\bm n}$ is the unit vector normal to the surface. For a cylinder of radius $R$ along $\hat{z}$ axis we have $\hat{n} = \cos{\varphi} \hat{x} + \sin{\varphi} \hat{y}$. The Hamiltonian that governs the TI surface modes in the presence of a sufficiently thin magnetic flux, $\phi = \eta \phi_0$, can then be written as

\begin{figure}[t]
\includegraphics[width=8cm]{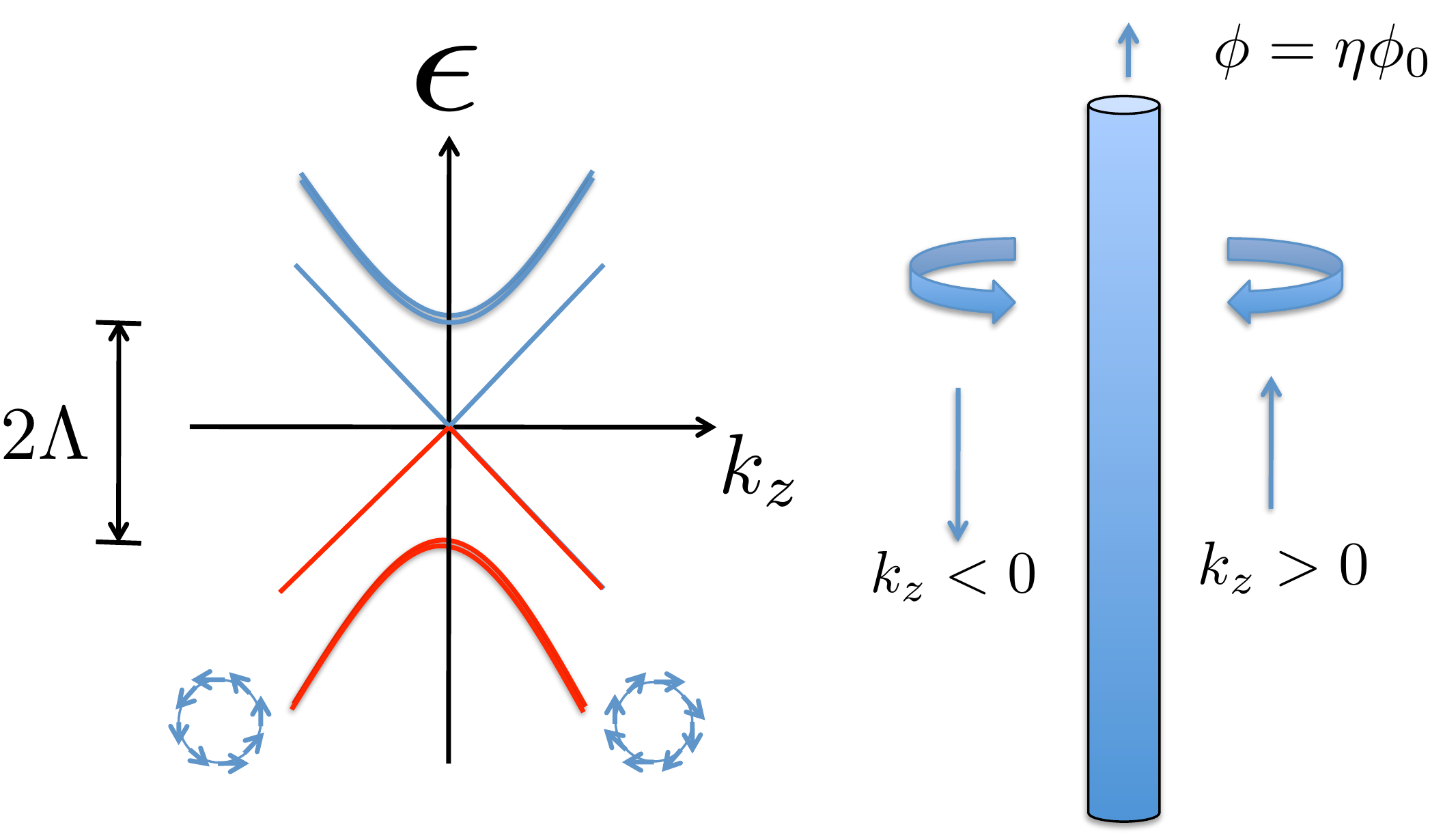}
\caption{(Color online) The surface spectrum of a wire made from a strong topological insulator in the presence of half integer multiple of the magnetic flux quantum along the wire. The bandwidth in which the spectrum is nondegenerate is $\Lambda = \hbar \upsilon_0 / R$. Where R is the radius of the wire. The spin texture is shown schematically for the states dispersing along $k_z$ with the blue arrows.}\label{fig1}
\end{figure}

\begin{equation}
H_k = \frac{1}{2 R} + \left(\begin{matrix} \frac{1}{R} (i \partial_{\varphi} + \eta) && -i k_z e^{- i \varphi} \\ i k_z e^{i \varphi} && -\frac{1}{R} (i \partial_{\varphi} + \eta)\end{matrix}\right),
\end{equation} 
\newline

We can simplify this Hamiltonian by a unitary transformation $\tilde{H}_k = U^{\dag}(\varphi) H_k U(\varphi)$ in which $U(\varphi)$ is given by 

\begin{equation}
U(\varphi) = \left( \begin{matrix} 1 && 0 \\ 0 && e^{i \varphi} \end{matrix} \right),
\end{equation}
using this transformation we can get rid of the phase in the off-diagonal components of the Hamiltonian matrix 

\begin{equation} 
H_k =\left(\begin{matrix} \frac{1}{R} (i \partial_{\varphi} + \eta - \frac{1}{2}) && -i k_z \\ i k_z  && -\frac{1}{R} (i \partial_{\varphi} + \eta -\frac{1}{2})\end{matrix}\right),
\end{equation}

The eigenstates of this Hamiltonian are given by $\psi_{k l}(\varphi) = e^{i \varphi l} \psi_{k l}$ where $ \psi_{k l}$ are eigenstates of the $H_{kl}$ defined as 

\begin{equation}
H_{k l} = k_z \sigma_2 - \frac{1}{R}\sigma_3 (l+ \frac{1}{2} - \eta),
\end{equation}
when there is a magnetic flux through the wire equal to the odd multiple of the half quantum of the magnetic flux $\eta = n + 0.5$ we get a non-degenerate gapless band for $l = n$

\begin{equation}
H_0 = k_z \sigma_2,
\end{equation}

We assume that $R$ is small enough in a way that there exist a significant energy range, $\Lambda$, in which the gapless band does not overlap with other bands (See Fig. \ref{fig1}) then we can use this energy band as a building block of the model to realize a Weyl semimetal phase.


\section{ Lattice of Parallel Wires}

\begin{figure}[t]
\includegraphics[width=5cm]{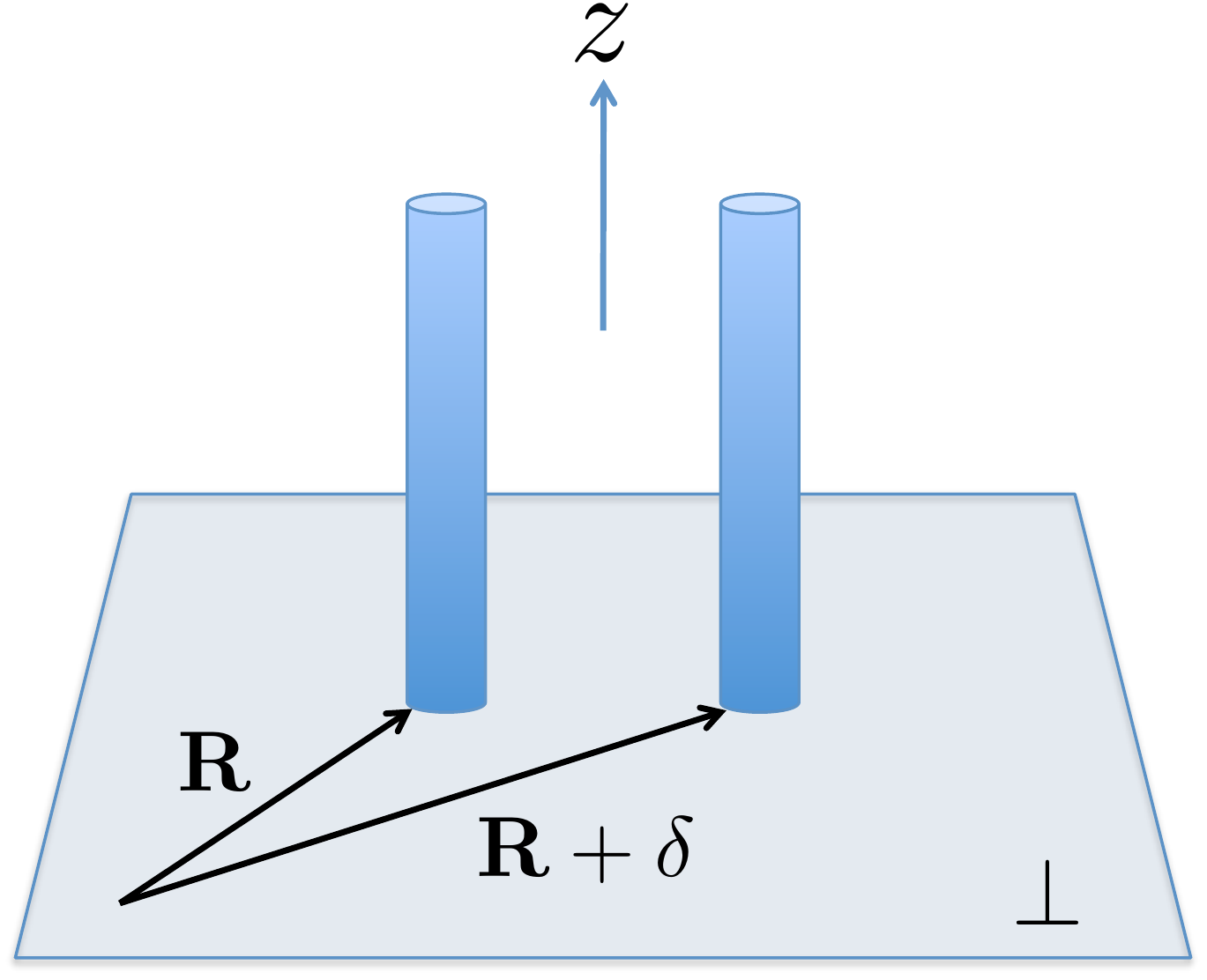}
\caption{(Color online) Two adjacent parallel wires. The electronic states on one can interact with those on the other. }\label{fig2}
\end{figure}

When the distance between two parallel nano wires is sufficiently small, there would be an overlap between electronic wave-functions and this leads to a hopping between adjacent electronic states of the surface modes.  Here we consider a honeycomb arrangement of these nano wires by considering the nondegenerate gapless modes of these wires that can be achieved by exposing them to a static magnetic field along the wires. We assume that the spin-momentum locking that happen at the surface of topological insulators are in opposite directions for wires in the A and B sublattices. We should point out here that all the by far discovered topological insulators happen to have the same direction for the spin-momentum locking. In theory, one can get a system with an opposite spin-momentum locking by simply changing the sign of the spin-orbit coupling required to get the original topological insulator system \cite{Vsign}. We also assume that the Dirac points for both A and B wires have the same energy, however, in the end we relax this assumption and we discuss how this is essential in order to get a Weyl semimetal phase. On the other hand, these wires are multi-band systems, therefore, they must be sufficiently thin to allow a significant gap, $\Lambda$, in which the gapless band does not overlap with other subbands (see Fig. (\ref{fig2})). With these assumptions and by considering only the non-degenerate lowest lying band in each wire in the energy range $2\Lambda$ (see Fig. \ref{fig1}) we can split the Hamiltonian into two parts as follows

\begin{equation}
H = H_0 + H_{\text t},
\end{equation}
in which $H_0$ which sums up the gapless nondegenarate modes of each individual wire in a second quantized notation and it is given in terms of the Fourier components as
\begin{equation}
H_0 = \int \frac{d k_z}{ 2 \pi} \sum_{\bk_{\perp}} \Psi^{\dag}_{\bk_{\perp}}(k_z) (\upsilon_0 k_z {\tau_3} \sigma_2) \Psi^{}_{\bk_{\perp}}(k_z),
\end{equation}
where $\tau_3$ is the pauli matrix acting on the sublattice Hilbert space, $k_z$ is the momentum along the wire (Fig. \ref{fig1}) and $\bk_{\perp}$ spans the honeycomb's reciprocal lattice (similar to what one gets for graphene) and $\Psi^{\dag} = (\psi^{\dag}_{A+},\psi^{\dag}_{A-},\psi^{\dag}_{B+},\psi^{\dag}_{B-})$. The first( last) two components act on the spin space of the wire in the A( B) sub-lattice. $\psi^{\dag}_{A/B\alpha}$ creates an electron in the $\alpha$ branch of the $A/B$ wire when it acts on the vacuum state.  Using this representation, $H_{\text t}$ which represents the direct hopping between 
nearest-neighbour wires, can be written as
\begin{equation}
H_{t} =  \int \frac{d k_z}{ 2 \pi} \sum_{\bk_{\perp}} \Psi^{\dag}_{\bk_{\perp}}(k_z) \left[ {\left(\begin{matrix} 0 && g(\bk_{\perp}) \\ \bar{g}(\bk_{\perp}) && 0 \end{matrix}\right)} \otimes {\mathbb 1} \right] \Psi^{}_{\bk_{\perp}}(k_z),
\end{equation}
in which $g(\bk_{\perp}) = -t \sum_{\delta_i} e^{-i \bk_{\perp} \cdot \delta_i}$ and $\delta_i$ are three vectors that connect a site in honeycomb lattice to the three adjacent points. The energy spectrum of the total Hamiltonian would then be doubly degenerate 

\begin{equation}
\varepsilon_{\pm}(k_z, \bk_{\perp}) = \pm \sqrt{\upsilon_0^2 k_z^2 + |g(\bk_{\perp})|^2},
\end{equation}
$g(\bk_{\perp})$ vanishes at two inequivalent points in the two-dimensional reciprocal lattice, i.e., at $\bK^{\pm}$ and is linear near these points.  Therefore, we get two degenerate three-dimensional Dirac points at $(k_z, \bk_{\perp}) = (0, \bK^{\pm})$. In order to get a topologically protected semimetal phase we need to break time-reversal or inversion symmetry in such a way that it separates the Dirac points. Breaking inversion symmety can be easily realized by relaxing the assumption we had in the beginning, i.e., Dirac point at two sub-lattices now can have different energies. By revising $H_0$ to account for such an energy difference, $V$, we get

\begin{equation}\label{EqInv}
H_0 = \int \frac{d k_z}{ 2 \pi} \sum_{\bk_{\perp}} \Psi^{\dag}_{\bk_{\perp}}(k_z) (\upsilon_0 k_z {\tau_3} \sigma_2 + V \tau_3) \Psi^{}_{\bk_{\perp}}(k_z),
\end{equation}

In this case the degeneracy is lifted and we have four bands $\varepsilon_{s r}(k_z, \bk_{\perp})$ ($s,r = \pm$) given by

\begin{equation}
\varepsilon_{s r}(k_z, \bk_{\perp}) = r \sqrt{(\upsilon_0 k_z + s V)^2 + |g(\bk_{\perp}) |^2}, 
\end{equation}

This spectrum has two pairs of Weyl points in the three dimensional BZ at which the bands cross linearly. Each pair consists of two opposite chirality Weyl points centred at $(k_z, \bk_{\perp}) = (0, \bK^{\pm})$. They are separated along $k_z$ axis by $Q = 2 V / \upsilon_0$. According to the argument presented in the previous section, these crossings are topologically protected against various local momentum conserving perturbations as long as they are separated in the BZ. It is important to note that although these isolated Weyl nodes are robust against local momentum-conserving perturbations, the peculiarity in the Hall response that is present in a system with only one pair of Weyl nodes does not exist here, i.e., the transverse conductivity in the presence of the applied magnetic field along the wires, $\sigma_{xy}$ is zero\cite{Haldane, weyl1} and two pairs have cancelling contributions to the Hall conductivity as they are separated in the opposite way at two valleys considering their chirality. There is also a possibility of phase transition to an insulator phase due to the electron-electron interactions. This instability arises in a mean-field treatment of the inter-rod electron-electron interactions. When the wire radius, R, is much smaller than the honeycomb lattice constant, $a$, the coulomb interaction between electrons in the adjacent wires can be written as 

\begin{equation}
H_{\text e-e} = U \int_{-L/2}^{L/2} d z  \int_{-\infty}^{\infty} d u \frac{\hat{n}(\bR,z) \hat{n}(\bR+ {\bm \delta},z+u)}{\sqrt{1+ (\frac{u}{a})^2}},
\end{equation}
where $\hat{n}$ is the electronic density operator and $U = e^2/(4 \pi \epsilon a)$. $\bR$, $\bm \delta$ and $z$ have been defined in the Fig. (\ref{fig2}) In a mean-field treatment of the above term and by considering a Kekule type modulating order parameter with a wave-vector $\bG = \bK^{+} - \bK^{-}$, it is possible to connect the Weyl points separated by $\bG$. This would then open up a gap and the system becomes an insulator. The critical coupling, $U_c$ at which such a phase transition to an insulator phase occurs can be computed for this system. It is a function of the potential difference, $V$, as well as the hoping strength, $t$, and is given by

\begin{figure}[t]
\includegraphics[width=8cm]{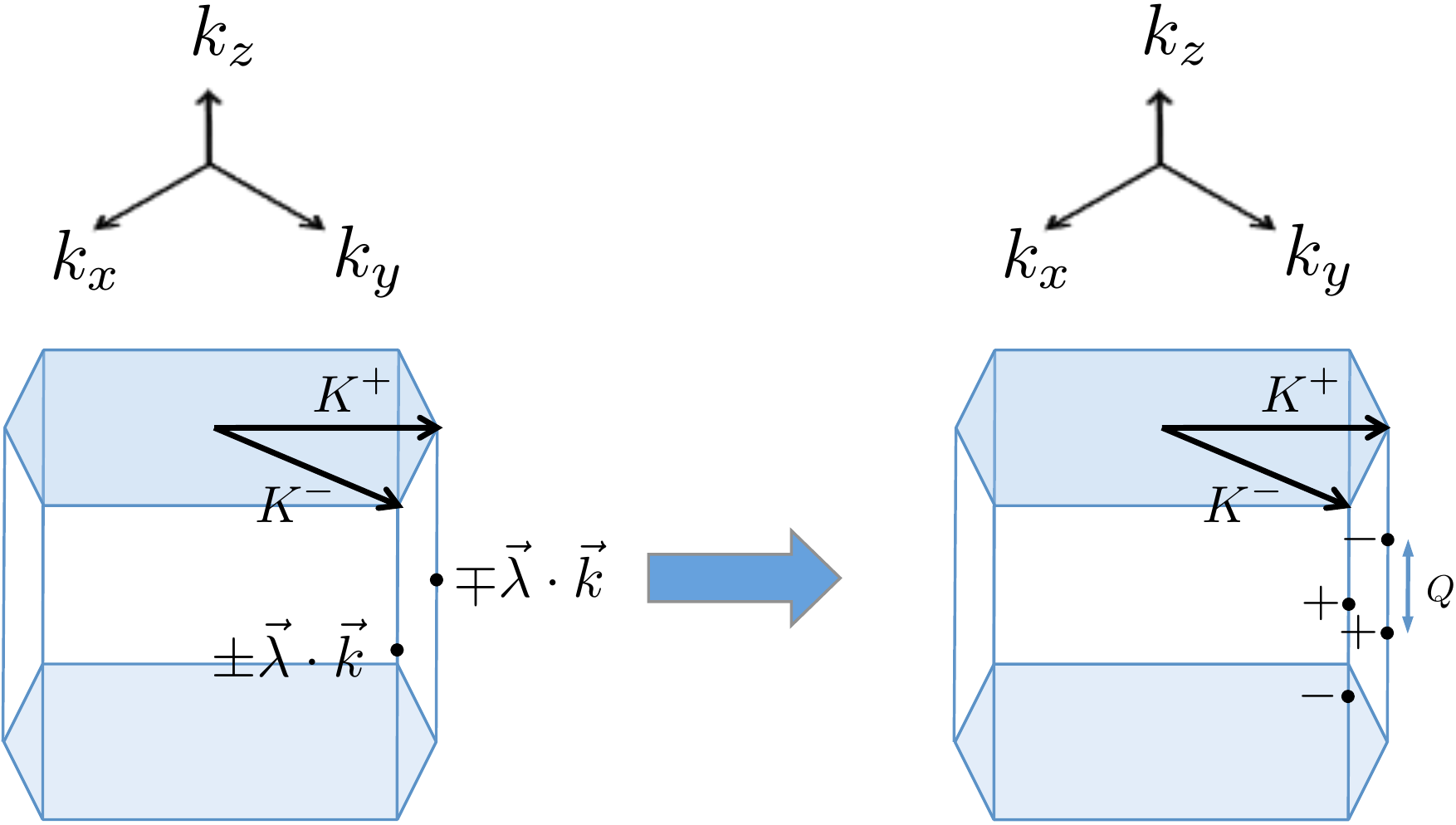}
\caption{(Color online) The three dimensional BZ before and after separating the Weyl nodes along the $k_z$. We have two inequivalent  pairs of opposite chirality Weyl nodes.}\label{fig3}
\end{figure}
\begin{figure}[t]
\includegraphics[width=6cm]{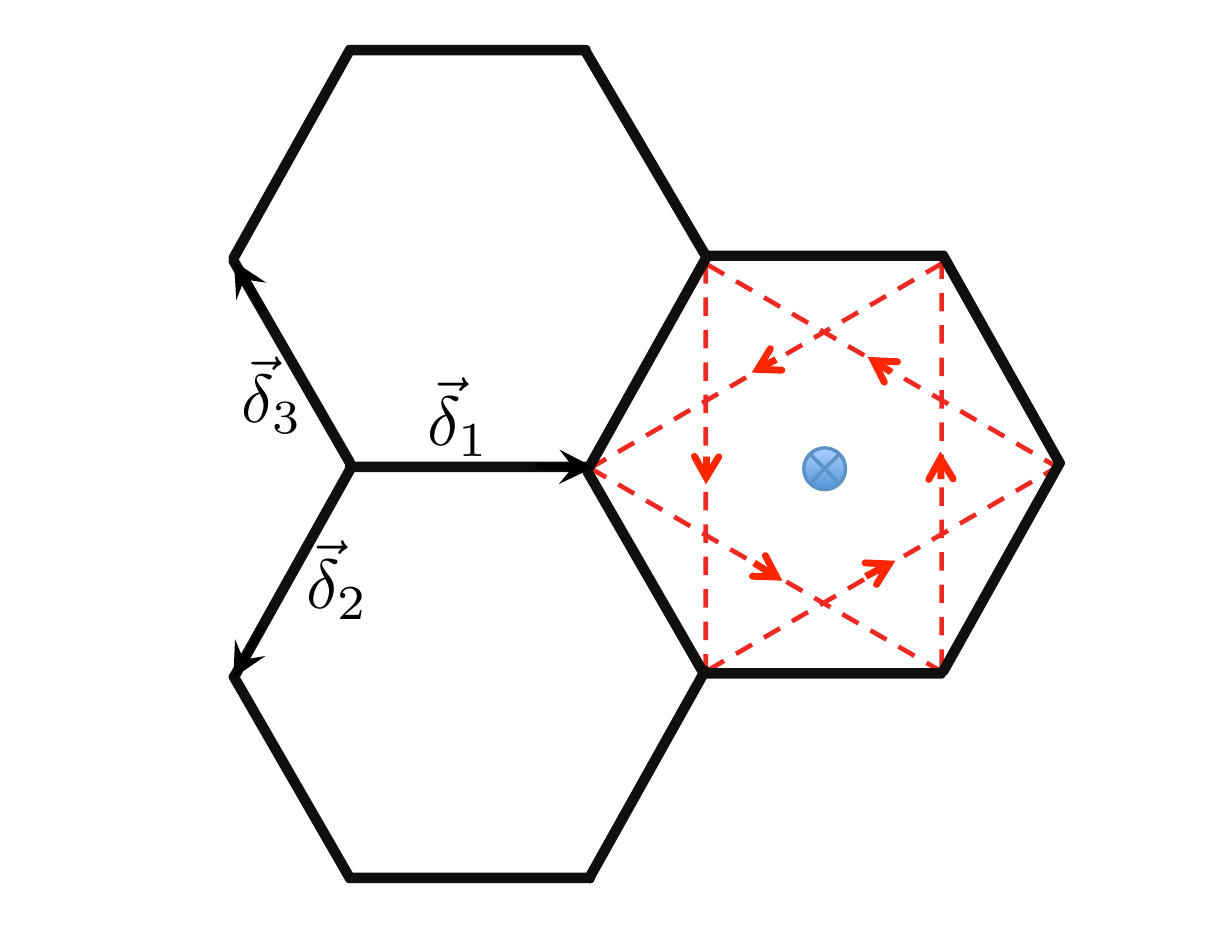}
\caption{(Color online) The modulations in the magnetic flux with a zero average through each hexagon introduces a complex next nearest neighbor hopping parameter, $t' e^{i \phi}$. The vectors that connect adjacent sites in the same sublattice are $\pm (\bb_1,\bb_2,\bb_3) =\pm ( {\bm \delta}_1 - {\bm \delta}_2, {\bm \delta}_2 - {\bm \delta}_3, {\bm \delta}_3 - {\bm \delta}_1$) }\label{fig4}
\end{figure}

\begin{equation}
U_c = \frac{4 \Lambda}{3} \left( 1 + \ln{\frac{\alpha V (\Lambda - V)}{t}}\right)^{-1},
\end{equation}
in which $\alpha = (8/\sqrt{3} \pi)^{1/2}$. This critical coupling would be of the order of the single wire's non-degenerate bandwidth $\Lambda$ (see Fig. \ref{fig1}) whenever the hoping strength is significant enough. For $U< U_c$ the system would remain in the topological semimetal phase and for $U>U_c$ the system would become a three-dimensional version of Kekule insulator which has been discussed previously in two dimensions for graphene\cite{Kekule}. It is possible to make the system stable against such phase transition for all ranges of interactions at least in this mean-field channel by introducing a term in the Hamiltonian which separates two Weyl nodes at each valley ($\bK^{+} $ and $\bK^{-}$) by a different amount. In this case the system becomes stable against the perturbation that connects two valleys since all the opposite chirality Weyl pairs would have incommensurate separation in momentum space after addition of such a term.  This term can be induced by considering next nearest-neighbour inter-wire hopping of the electronic states. The hopping amplitude is imaginary considering a modulating magnetic flux with a zero net average through each hexagon (see Fig. 4). This can be achieved by applying a modulating magnetic field instead of the uniform magnetic field that is required to induce half-quantum magnetic fluxes in the wires. Such a term was first introduced for a honeycomb lattice to realize a quantum Hall phase in a system without a net uniform applied magnetic field\cite{Haldane}.  The Hamiltonian for this next-nearest neighbor hopping in the momentum space can be written as

\begin{equation}
H' = -t' \int \frac{d k_z}{ 2 \pi} \sum_{\bk_{\perp}} \Psi^{\dag}_{\bk_{\perp}}(k_z) ( \mu(\bk_{\perp} ) + V(\bk_{\perp}) \tau_3  ) \Psi^{}_{\bk_{\perp}}(k_z),
\end{equation}
which is similar to the term which has been introduced in Eq. (\ref{EqInv}) but here the $V$ which separates the Dirac nodes along $k_z$ is a function of the $\bk_{\perp}$ and takes different values at $\bK^{+}$ and $\bK^{-}$. The presence of $\mu(\bk_{\perp})$ separates two pairs in energy since it also depends on $\bk_{\perp}$ and can take different values at two valleys. In the original Haldane's \cite{Haldane} flux configuration $\mu(\bk_{\perp})$ and $V(\bk_{\perp})$ are given by 

\begin{equation}\label{VH}
V(\bk_{\perp}) = -2 t' \sin{\phi} \sum_{i=1,2,3} \sin{\bk_{\perp} \cdot {\bm b}_i},
\end{equation}
and 
\begin{equation}\label{MuH}
\mu(\bk_{\perp}) = 2 t' \cos{\phi} \sum_{i=1,2,3} \cos{\bk_{\perp} \cdot {\bm b}_i},
\end{equation}
${\bm b}_i$ and $\phi$ have been defined in the caption of Fig. 4 and $t'$ is the next-nearest neighbor hopping amplitude.
 
Therefore, the distance between each pair of Weyl nodes is now an incommensurate wave-vector since the separations of Weyl nodes at two valleys, i.e., $\bq_1 = 2V(\bK^{+}) \hat{z}/\upsilon_0$ and $\bq_2 =  2V(\bK^{-})\hat{z}/\upsilon_0 $ are not exactly opposite. This makes the system stable against charge density perturbations that connect quantum states of two valleys. The Axion field is not zero in this case as the separations are not exactly opposite and instead it modulates in space\cite{anomaly4}.

Therefore in the presence of the Haldane term the system breaks the time reversal symmetry and it would have a nontrivial topological response. In this case the transverse conductivity in the presence of the magnetic field in the $z$ direction can be obtained by a summation over the transverse conductivity of the massive two-dimensional Dirac fermions for each $k_z$\cite{Haldane, weyl1} . At $\bk_{\perp} = \bK^{+}$ and $-q_1<k_z<q_1$ the sign of the masses of two-dimensional Dirac states are in such a way that they contribute constructively to the transverse Hall conductivity. Similarly, at $\bk_{\perp} = \bK^{-}$ and $-q_2<k_z<q_2$ the transverse Hall conductivity is nonzero. The net $\sigma_{xy}$ can be obtained, when the chemical potential is inside the gap in the presence of an applied magnetic field in the $z$ direction, by a summation over the contribution of all two-dimensional Dirac fermions labeled by $k_z$ and the valley index, i.e., 1 and 2

\begin{equation}
\sigma_{xy} = \int_{-\Lambda}^{\Lambda} [\nu_1({k_z}) - \nu_2({k_z})]\frac{d k_z}{2\pi} \frac{e^{2}}{h},  
\end{equation}
in which $\nu_1 = \Theta (q_1 - |k_z|)$ and $\Theta(q_2 - |k_z|)$. Using Eq. (\ref{VH}) the transverse conductivity becomes 

\begin{equation}
\sigma_{xy} = \frac{6 \sqrt{3} t'  \sin{\phi} \;e^{2}}{\pi h},
\end{equation}
\\

This result highlights the fact that although the momentum conserving perturbations cannot gap out the system, this does not necessarily imply the existence of a topological electromagnetic response in the system and two pairs of Weyl nodes can have cancelling contribution to the electromagnetic response in special cases and therefore $\sigma_{xy}$ can be zero even when the Weyl nodes are well separated in the BZ which is the case for our model in the absence of the Haldane term. Adding the next nearest hoping and introducing the Haldane term would lead to a nonzero net topological electromagnetic response since two Weyl pairs would now have different separations and therefore their contributions to the transverse conductivity do not sum up to zero.

\section{Conclusions}

To summarize we have considered a system of parallel nano wires in the time-reversal invariant topological insulator phase and by breaking inversion and time-reversal symmetry we found that it is possible to realize a topological semimetal phase with two pairs of Weyl nodes in the three dimensional BZ which are protected against local perturbations that conserve crystal momentum. A nontrivial topological electromagnetic response might arise under certain circumstances when the contributions from two pairs does not exactly cancel each other. It would also be interesting to study the surface modes in the presence of more than one pairs of Weyl nodes which can be done using the model introduced in this paper. 

Finally, we point out that although the growths of large-scale vertically aligned nano-rods and nanopillars with various arrangements have already been achieved by experimentalists \cite{expGrow1, expGrow2, expGrow3}, realizing a system for which one can observe the robust Weyl energy crossings seems not to be experimentally plausible at this point. One can name various obstacles on the way of its experimental realization. First, wires are multi-band systems and the two-band model approximation used throughout this paper requires a significant gap for the other existing subbands in the nano-wires. For the so far discovered topological insulators, the surface Fermi velocity\cite{hasan2}, $\upsilon_0$, is of the order of $10^5 m/s$ ($5 \times 10^5 m/s$ for Bi$_2$Se$_3$ \cite{shen1}), therefore, the wire diameter required ($2R \lesssim 60$nm) to get a significant gap ($\Lambda\gtrsim10$meV) is still out of the experimentally feasible sub-micron ranges\cite{expGrow1, expGrow2, expGrow3}. Finally, another challenge to realize such a system is to find two types of topological insulators with opposite spin-momentum lockings which although in theory nothing prevents having such systems, in all the so far discovered topological insulators surface electrons' spins curl around the Dirac point in the momentum space in the same directions. We note that the direction of the spin-momentum locking is a material dependent property and in the theoretical lattice models for the topological insulators in three dimensions, it is possible to change it on the surface by varying physical parameters of the bulk\cite{Vsign}. 

\section{Acknowledgment}
The author is much indebted to Professor Marcel Franz for his illuminative comments and support during the course of this work. The author has also benefited from a fruitful discussion 
with Professor Allan H. MacDonald during the meeting of the Cifar quantum material program in Toronto, May 2012.



\begin{thebibliography}{10}
\bibitem{Vsign}
Another way of thinking how one can achieve this is to consider having 1D worm holes instead of rods in the bulk of a TI. In this situation since the direction of the surface normal vector (see Eq. (\ref{hcurve})) is opposite to the one defined for a TI rod the spin-momentum locking would be in the opposite direction.
\end{thebibliography}
\end{document}